# Transformation of a graphene nanoribbon into a hybrid 1D nanoobject with alternating double chains and polycyclic regions

Alexander S. Sinitsa[a], Irina V. Lebedeva[b], Yulia G. Polynskaya[c], Dimas G. de Oteyza[d,e,f], Sergey V. Ratkevich[g], Andrey A. Knizhnik[c], Andrey M. Popov[h†], Nikolai A. Poklonski[g] and Yurii E. Lozovik[h,i]

[a.] National Research Centre "Kurchatov Institute", Kurchatov Square 1, Moscow 123182, Russia.
[b.] CIC nanoGUNE BRTA, Avenida de Tolosa 76, San Sebastian 20018, Spain.
[c.] Kintech Lab Ltd., 3rd Khoroshevskaya Street 12, Moscow 123298, Russia.
[d.] Donostia International Physics Center, San Sebastián, Spain.
[e.] Centro de Física de Materiales (CFM-MPC), CSIC-UPV/EHU, San Sebastián, Spain.
[f.] Ikerbasque, Basque Foundation for Science, Bilbao, Spain.
[g.] Physics Department, Belarusian State University, Nezavisimosti Ave. 4, Minsk, 220030, Belarus.
[h.] Institute for Spectroscopy of Russian Academy of Sciences, Fizicheskaya Street 5, Troitsk, Moscow 108840, Russia.
[i.] National Research University Higher School of Economics, 109028 Moscow, Russia.

**ABSTRACT:** Molecular dynamics simulations show that a graphene nanoribbon with alternating regions which are one and three hexagons wide can transform into a hybrid 1D nanoobject with alternating double chains and polycyclic regions under electron irradiation in HRTEM. A scheme of synthesis of such a nanoribbon using Ullmann coupling and dehydrogenation reactions is proposed. The reactive REBO-1990EVC potential is adapted for simulations of carbon-hydrogen systems and is used in combination with the CompuTEM algorithm for modeling of electron irradiation effects. The atomistic mechanism of formation of the new hybrid 1D nanoobject is found to be the following. Firstly hydrogen is removed by electron impacts. Then spontaneous breaking of bonds between carbon atoms leads to decomposition of narrow regions of the graphene nanoribbon into double chains. Simultaneously thermally activated growth of polycyclic regions occurs. Density functional theory calculations give barriers along the growth path of polycyclic regions consistent with this mechanism. Electronic properties of the new 1D nanoobject are shown to be strongly affected by the edge magnetism and make this nanostructure promising nanoelectronic and spintronic applications. The way of synthesis the 1D nanoobject proposed here can be considered as an example of the general three-stage strategy of production of nanoobjects and macromolecules. 1) precursors are synthesized by traditional chemical method, 2) precursors are placed in HRTEM with the electron energy that is sufficient only to remove hydrogen atoms, 3) as a result of hydrogen removal, the precursors become unstable or metastable and transform into new nanoobjects or macromolecules.

### 1. Introduction

The progress in high-resolution transmission electron microscopy (HRTEM) has made it possible to study transformations of nanostructures induced by electron irradiation *in situ*. In particular, notable observations for carbon nanostructures include the reconstruction of zigzag graphene edges terminated by hexagons into alternating pentagons and heptagons[1,2], the evolution of atomic-scale vacancies in the graphene layer into holes[3,4] and the cutting of carbon nanotubes by osmium[5] and nickel[6] clusters under the simultaneous action of electron irradiation. Structural transformations of carbon nanostructures induced by electron irradiation in HRTEM give rise to formation of nanoobjects initially absent in the sample. Examples of such nanoobject formation are the transformation of a graphene flake into a fullerene[7], the formation of a double-wall carbon nanotube from a peapod carbon nanotube[8], the transformation of a bilayer graphene nanoribbon (GNR) into a flat carbon nanotube[9], or the transformation of GNRs[1,9—12] and carbon nanotubes[13,14] into a single, double and triple atomic carbon chains. Moreover, transformation under electron irradiation can lead to synthesis of nanoobjects which have not been obtained by other methods. For instance, formation of an endohedral metallofullerene from a nickel cluster surrounded by amorphous carbon has been observed in HRTEM[15], whereas endohedral metallofullerenes with transition metal core have not been produced yet by traditional methods of fullerene synthesis.

Although HRTEM is a useful tool for the formation of new nanoobjects, their synthesis is limited by the availability of appropriate initial nanostructures that may be transformed under electron irradiation into these nanoobjects. For example, formation of long atomic carbon chains of several hundred atoms in length upon heating of zigzag graphene nanoribbons (GNR) with regular structure of two hexagon rows has been predicted by recent molecular dynamics (MD) simulations[16], while the typical length of chains obtained in HRTEM from GNRs with irregular structure is only tens of atoms[1,9—12]. Thus, GNRs can be used as a starting point for formation of new 1D nanoobjects under electron irradiation. Early works on GNR production typically relied on top-down production methods (lithography, unzipping of carbon nanotubes…) that unfortunately lack atomic precision. However, in 2010 the seminal work of Cai et al. demonstrated that atomically precise GNRs can be synthesized by a bottom-up approach[17]. Appropriate precursor molecules were dosed onto a surface and heated to activate their chemical transformation. Their chemical design was such that the activated reactions followed well-defined paths, ultimately resulting in GNRs whose structure was precisely determined by the utilized reactant. Such findings triggered enormous research efforts along these lines that have led to great advances, as is the successful synthesis of atomically precise GNRs with different widths, edge orientations, edge structures, heteroatoms, or with periodically added side groups[18,19]. Most of those achievements are based on adequately modified designs of the reactant molecules, and further variations may still provide many other GNRs with new structures to be investigated in the future. By way of example, the use of appropriately functionalized polyacene derivatives may result, after Ullmann coupling and dehydrogenation reactions[20], in a new type of GNR with alternating regions of different widths of one and three hexagon rows, as described below. Hydrogen can be easily removed by electron irradiation in HRTEM. In turn, the one-hexagon wide zigzag GNR segment is unstable and decomposes into two chains[10] once it loses its hydrogen terminations. We thus suggest that electron irradiation of the proposed GNR can lead to formation of a hybrid 1D nanoobject with alternating chains and polycyclic graphene-like regions.

To demonstrate the possibility of obtaining the new 1D carbon nanoobjects, we have performed MD simulations of the transformation process under the action of electron irradiation using the CompuTEM algorithm[21,22]. This algorithm takes into account the annealing of the system between changes induced by irradiation in the local structure and allows to rescale the structure evolution time to the experimental conditions in HRTEM taking into account electron scattering cross-sections. Previously, MD simulations using the CompuTEM algorithm have been applied to study transformation of graphene flake into fullerene[21,22], formation of metal heterofullerenes[23] and endohedral metallofullerenes[15] and cutting of carbon nanotubes by nickel cluster[6] under electron irradiation. The excellent agreement of principal features of structural transformations and structure evolution times with the experimental HRTEM observations have been revealed[6,15].

Recently we reparametrized the popular Brenner potential[24] for carbon-hydrogen systems to describe energies of pristine graphene edges, vacancy migration in graphene and formation of atomic carbon chains[16,25]. In the present paper we test the potential for graphene edges functionalized by hydrogen and modify the relevant parameters to improve its performance. This potential, specifically suited for modelling of transformations of carbon nanostructures, is used in our MD simulations.

To distinguish the roles of spontaneous and irradiation-induced breaking and formation of bonds between carbon atoms, electron beams with different kinetic energies of electrons are considered here in the MD simulations. The atomistic mechanism of formation of 1D nanoobjects and fractions of polycyclic graphene-like regions with different structure derived from the wide regions of the GNR under electron irradiation are determined. The MD simulations are supplemented with DFT calculations of barriers for formation of new bonds between carbon atoms to confirm the atomistic mechanism of formation of the 1D nanoobjects revealed. The electronic and magnetic properties of the 1D nanoobject with the most abundant polycyclic regions are also studied by the DFT calculations.

The paper is organized in the following way. Section 2 is devoted to the re-parametrization of the REBO potential used and details of MD simulations and DFT calculations. The results of MD simulations of formation of hybrid 1D nanoobjects as well as DFT calculations of their structure, electronic properties and the energetics along the growth path are presented in Section 3. The discussion and conclusions are given in Section 4.

## 2. Methods
### 2.1. Modification of interatomic potential

The original Brenner potential[24] (REBO-1990) was designed to describe the bulk of carbon crystals and small hydrocarbons. In the present paper, however, we perform simulations for GNRs, which are different from both of these cases. It is important for us to describe properly not only the graphene bulk but also its edges. Irradiation of GNRs implies formation of local structure defects such as vacancies as well as atomic chains at the edges. To adequately model pristine graphene edges, vacancy migration in graphene and formation of atomic carbon chains, we have previously reparametrized the Brenner potential[16,25] (REBO-1990EVC). Here we test this potential for graphene edges functionalized by hydrogen and improve some of the corresponding parameters (REBO-1990EVC_CH).

In REBO-1990, the binding energy is written in the form

$$E_b = \sum_i \sum_{j>i} \left( V_R(r_{ij}) - \bar{B}_{ij} V_A(r_{ij}) \right) \quad (1)$$

where $V_R(r_{ij})$ and $V_A(r_{ij})$ are repulsive and attractive pairwise terms for atoms $i$ and $j$ with the distance $r_{ij}$ between them and $\bar{B}_{ij}$ is the bond-order term describing how the bond energy depends on the coordination.

The bond-order term is computed as

$$\bar{B}_{ij} = \left( B_{ij} + B_{ji} \right)/2 + F\left(N_{ij}^t, N_{ji}^t, N_{ij}^{conj}\right)/2 \quad (2)$$

where $B_{ij}$ and $B_{ji}$ are the bond orders for atoms $i$ and $j$ and $F$ is the correction function introduced to distinguish radicals and the difference between conjugated and non-conjugated bonds. The function $F$ depends on the total numbers of neighbours $N_{ij}^t$ and $N_{ji}^t$ of atoms $i$ and $j$ excluding themselves and on $N_{ij}^{conj}$ used to check whether the bond is the part of the conjugated system or not. The bond order is given by

$$B_{ij} = \left(1 + \sum_{k \neq i,j} A_{ijk} + H_{ij}\left(N_{ij}^{(H)}, N_{ij}^{(C)}\right)\right)^{-\delta_i} \quad (3)$$

where $A_{ijk}$ corresponds to the bond screening by atoms $k$ in the local chemical environment of atom $i$ in solid carbon phases and the function $H_{ij}$ is a correction to the solid-state bond order required to accurately describe molecular bond energies. The function $H_{ij}$ depends on the number of hydrogen and carbon neighbours of atom $i$, $N_{ij}^{(H)}$ and $N_{ij}^{(C)}$, respectively.

The total number of neighbours is found as

$$N_{ij}^t = \sum_{k \neq i,j} f(r_{ik}) \quad (4)$$

where $f(r)$ is the cutoff function, which changes from 1 to 0 upon increasing the distance $r$:

$$f(r) = \begin{cases} 1, & r < R^{(1)} \\ \frac{1}{2}\left(1 + \cos\frac{\pi(r - R^{(1)})}{(R^{(2)} - R^{(1)})}\right), & R^{(1)} \leq r \leq R^{(2)} \\ 0, & r > R^{(2)} \end{cases} \quad (5)$$

The numbers $N_{ij}^{(H)}$ and $N_{ij}^{(C)}$ are defined in the way similar to $N_{ij}^t$ (see eq. (4)) but summation is performed only over hydrogen or carbon neighbours of atom $i$ except atom $j$, respectively.

The number $N_{ij}^{conj}$ is expressed as

$$N_{ij}^{conj} = 1 + \sum_{k \neq i,j} f(r_{ik})\Phi(N_{ki}^t) + \sum_{k \neq i,j} f(r_{jk})\Phi(N_{kj}^t) \quad (6)$$

where $\Phi(x)$ is 1 for $x \leq 2$ and 0 for $x \geq 3$:

$$\Phi(x) = \begin{cases} 1, & x \leq 2 \\ (1 + \cos\pi(x-2))/2, & 2 < x < 3 \\ 0, & x \geq 3 \end{cases} \quad (7)$$

The fitting set for the pairwise terms and bond orders of REBO-1990 included the experimental and first-principles results on cohesive energies and lattice constants of diverse carbon crystals: diamond, graphite, simple cubic and face-centered cubic structures. The function $F(n_1, n_2, n_3)$ for integer numbers $n_1$, $n_2$ and $n_3$ was obtained from atomization energies of small hydrocarbons and tight-binding data on the formation energies of vacancies in graphite and diamond. For non-integer numbers $n_1$, $n_2$ and $n_3$ the values of the function $F$ are found by interpolation. Also the function $F$ complies with the conditions $F(n_1, n_2, n_3) = F(n_2, n_1, n_3)$ and $F(n_1, n_2, n_3 > 2) = F(n_2, n_1, 2)$. Two sets of the parameters were presented in the original Brenner paper but we use only the second set (Table III of Ref. 24), which provides better force constants.

In our previous version REBO-1990EVC of the potential[16,25], the following parameters were changed: (1) $F(1,1,2)$ to fit the formation energy of carbon chains; (2) $F(1,2,2)$ to fit the energies of pristine $ZZ_0$ (zigzag), $ZZ(57)_{00}$ (reconstructed zigzag with alternating pentagons and heptagons[26]) and (armchair) edges; (3) $F(2,3,2)$ to fit the energy of the symmetric saddle point for vacancy migration. Here we check the performance of this potential for various edges functionalized by hydrogen: $AC_{11}$, $AC_{21}$, $AC_{22}$, $ZZ(57)_{11}$, $ZZ_1$, $ZZ_{211}$, $ZZ_{221}$ and $ZZ_2$ (Figure 1, the subscript denotes the number of hydrogen atoms bonded to each edge carbon atom within the elementary period of the GNR[27]). We propose some improvements within the new version REBO-1990EVC_CH. The parameters of the different versions of the potential are listed in Table 1.

**Table 1. Modified parameters of different versions of the Brenner potential.**

| Parameter | REBO-1990 | REBO-1990EVC | REBO-1990EVC_CH |
|---|---|---|---|
| $F(1,1,2)$ | 0.108 | 0.02514818 | 0.02514818 |
| $F(1,2,2)$ | −0.0243 | −0.038 | −0.038 |
| $F(2,3,2)$ | −0.0363 | −0.088 | −0.088 |
| $F(3,3,2)$ | 0 | 0 | −0.156 |
| $H_{CH}(1,2)$ | −0.4449 | −0.4449 | −0.615 |

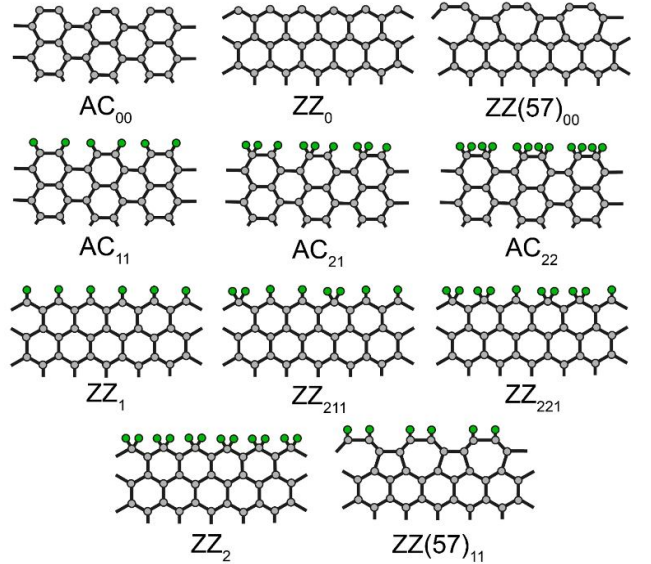

**Figure 1.** Structures of graphene edges functionalized with hydrogen. Carbon and hydrogen atoms are coloured in grey and green, respectively. AC, ZZ, and ZZ(57) are armchair, zigzag, and reconstructed zigzag edges, respectively. The subscripts indicate the number of hydrogen atoms bonded to each carbon atom at the edge within the elementary period.

To compute the edge energies, we consider 20-AGNR and 36-ZGNR (armchair GNR with 20 rows and zigzag GNR with 36 rows, respectively, see paper [28] for GNR notations). We start from the structures with the bond length equal to the optimal one for the infinite graphene layer. The simulation box with periodic boundary conditions includes 18 and 10 unit cells along the nanoribbon axis for the AGNR and ZGNR, respectively. To obtain the ribbon with the $ZZ(57)_{00}$ edges, the hexagons at the 36-ZGNR edges are converted into alternating pentagons and heptagons. To construct hydrogen-functionalized nanoribbons, hydrogen atoms are added at the distance of 1 Å from the edge carbon atoms. The structures of the GNRs are geometrically optimized by the method of conjugated gradients keeping the size of the simulation box fixed till the energy change in consecutive iterations becomes less than $10^{-10}$ eV/atom. The edge energies per unit edge length are found as $E_{ed} = (E_{GNR} - N_C \varepsilon_{gr} - N_H \varepsilon_H)/(2L_{GNR})$, where $E_{GNR}$ is the ribbon energy per unit cell, $N_C$ and $N_H$ are the numbers of carbon and hydrogen atoms in the ribbon per unit cell, respectively, $\varepsilon_{gr}$ is the binding energy of bulk graphene per atom, $\varepsilon_H$ is the binding energy of hydrogen molecule per atom and $L_{GNR}$ is the length of the GNR unit cell. For all the considered versions of the potential, $\varepsilon_{gr} = -7.376$ eV and $\varepsilon_H = -2.375$ eV. To test the performance of different versions of the potential for small hydrocarbons, we have also performed geometrical optimization for the molecules considered in the original Brenner paper[24] with the stopping criterion given above.

The energies of pristine and functionalized graphene edges computed with different versions of the Brenner potential (Table 1) are presented in Table 2. For reference, available DFT data[26-32] are also given. A more detailed comparison with the results of DFT calculations for pristine edges can be found in Ref. 16.

Original REBO-1990 describes very well (within 0.1 eV/Å from the most common DFT results) the energies of the $ZZ(57)_{00}$, $AC_{00}$, $ZZ(57)_{11}$, $AC_{11}$, $AC_{21}$ and $ZZ_1$ edges (Table 2). However, the energy of the $ZZ_0$ is underestimated by 0.2–0.3 eV/Å. As follows from the DFT calculations[26,27,29,32], the $ZZ_1$ edge should be 0.05–0.08 eV/Å less stable than the $AC_{11}$ edge, while according to REBO-1990, the $ZZ_1$ edge is by very little but more stable than the $AC_{11}$ edge (just by 0.01 eV/Å). The DFT calculations[27] also predict that the energy of the zigzag edge should depend non-monotonically on the fraction of edge carbon atoms bonded to two hydrogen atoms. The smallest edge energy corresponds to $ZZ_{211}$, which is more stable than $ZZ_1$ and $ZZ_{221}$ by about 0.1 eV/Å and than $ZZ_2$ by 0.2 eV/Å. Differently, REBO-1990 predicts a monotonic growth of energy from $ZZ_1$ to $ZZ_2$ with an energy increase by about 0.08 eV/Å upon increasing the fraction of edge carbon atoms bonded to two hydrogen atoms by one third. REBO-1990 also overestimates by more than 0.2 eV/Å the energy of the $AC_{22}$ edge. This edge should be more favourable than the $AC_{11}$ edge[27,32] but this is not the case for REBO-1990.

In REBO-1990EVC, the energy of the pristine $ZZ_0$ edge is improved (Table 2). However, the energies of the edges in which each edge carbon atom is bonded to one hydrogen atom are the same as in REBO-1990 and the energies of the edges where there are carbon atoms bonded to two hydrogen atoms are worse. The energy of the edges where each edge carbon atom is bonded to one hydrogen atom, $ZZ_1$, $AC_{11}$ and $ZZ(57)_{11}$, can be modified through the parameters $H_{CH}(0,2)$ and $F(2,2,2)$. Nevertheless, the energy of the bonds inside graphene layers also depends on $F(2,2,2)$, which is set strictly to zero, and changing this parameter would destroy description of bulk graphene properties[24]. Modifying only $H_{CH}(0,2)$ we do not achieve to correct the relative stability of the $ZZ_1$ and $AC_{11}$ edges, while their absolute energies get even worse. Therefore, there is no point in changing this parameter either and we leave the description of the $ZZ_1$, $AC_{11}$ and $ZZ(57)_{11}$ edges the same as in the original version of the potential.

**Table 2. Energies of pristine and hydrogen-terminated edges (in eV/Å) calculated using different versions of the Brenner potential as compared to the DFT data from literature.**

| Edge | REBO-1990 | REBO-1990EVC | REBO-1990EVC_CH | DFT |
|---|---|---|---|---|
| | | Pristine | | |
| $AC_{00}$ | 0.999 | 1.032 | 1.032 | 1.0078[a], 1.202[b], 0.98[c] |
| $ZZ_0$ | 1.032 | 1.198 | 1.198 | 1.1452[a], 1.391[b], 1.31[c], 1.21[d] |
| $ZZ(57)_{00}$ | 0.937 | 0.966 | 0.966 | 0.9650[a], 1.147[b], 0.96[c], 0.97[d] |
| | | H-terminated | | |
| $AC_{11}$ | 0.078 | 0.078 | 0.078 | 0.0321[a], 0.012[b], 0.01[c], 0.019[f] |
| $AC_{21}$ | 0.231 | 0.487 | 0.096 | 0.2092[a] |
| $AC_{22}$ | 0.153 | 0.405 | -0.070 | -0.0710[a], -0.186[e], -0.174[f] |
| $ZZ_1$ | 0.067 | 0.067 | 0.067 | 0.0809[a], 0.090[b], 0.06[c], 0.11[d], 0.105[e], 0.100[f] |
| $ZZ_{211}$ | 0.154 | 0.301 | 0.077 | 0.0119[a], 0.03[d], -0.016[e], -0.011[f] |
| $ZZ_{221}$ | 0.237 | 0.530 | 0.079 | 0.1007[a] |
| $ZZ_2$ | 0.321 | 0.759 | 0.080 | 0.2224[a] |
| $ZZ(57)_1$ | 0.293 | 0.293 | 0.293 | 0.3337[a], 0.352[b], 0.34[c], 0.36[d] |

[a] Ref. 27 (PBE); [b] Ref. 29 (LDA); [c] Ref. 26 (PBE); [d] Ref. 30 (PW91); [e] Ref. 31 (LDA); [f] Ref. 32 (LDA)

The energies of the zigzag edges where there are carbon atoms bonded to two hydrogen atoms, $ZZ_{211}$, $ZZ_{221}$ and $ZZ_2$, depend on the parameters $H_{CH}(1,2)$ and $F(2,3,2)$. The latter was fitted in REBO-1990EVC to describe the structure of the symmetric transition state and the corresponding barrier for vacancy migration[16] and it is undesirable to change it again. It is not possible to describe properly the non-monotonic dependence of the edge energy on the fraction of edge carbon atoms bonded to two hydrogen atoms changing only one parameter $H_{CH}(1,2)$. However, modifying this parameter in REBO-1990EVC_CH (Table 1), we at least achieve that the energies of the $ZZ_{211}$, $ZZ_{221}$ and $ZZ_2$ edges become close to the energy of the $ZZ_1$ edge and deviate by no more than 0.15 eV/Å from the results of the DFT calculations[27] (Table 2). The energies of the armchair edges where there are carbon atoms bonded to two hydrogen atoms, $AC_{21}$ and $AC_{22}$, depend additionally on $F(3,3,2)$. We tune this parameter (Table 1) to fit the energy of the $AC_{22}$ edge and to provide the correct order of the armchair edges functionalized by hydrogen in energy (Table 2).

Modification of several parameters in REBO-1990EVC and REBO-1990EVC_CH as compared to REBO-1990 deteriorates the description of the energies of some small hydrocarbons (Table 3). But such species are not important for our simulations. Both pristine and hydrogen-terminated edges relevant for our simulations are sufficiently well described in REBO-1990EVC_CH (Table 2). Furthermore, it properly describes vacancy migration and chain formation, the same as the previous version REBO-1990EVC[16,25]. Therefore, in the following we use REBO-1990EVC_CH in our studies.

**Table 3. Atomization energies of hydrocarbons (in eV) calculated using different versions of the Brenner potential in comparison with the reference experimental data from Ref. 24.**

| Hydrocarbon | REBO-1990 | REBO-1990EVC | REBO-1990EVC_CH | Experiment |
|---|---|---|---|---|
| Alkanes | | | | |
| methane | 17.6 | 17.6 | 17.6 | 17.6 |
| ethane | 29.7 | 29.7 | 29.7 | 29.7 |
| propane | 42.0 | 42.0 | 43.7 | 42.0 |
| n-butane | 54.3 | 54.3 | 57.7 | 54.3 |
| n-pentane | 66.5 | 66.5 | 71.6 | 66.6 |
| cyclopropane | 35.0 | 35.0 | 41.0 | 35.8 |
| cyclobutane | 47.8 | 47.8 | 52.5 | 48.2 |
| cyclohexane | 73.6 | 73.6 | 83.8 | 73.6 |
| Alkenes | | | | |
| ethylene | 23.6 | 23.6 | 23.6 | 23.6 |
| propene | 36.2 | 35.7 | 35.7 | 36.0 |
| 1-butene | 48.5 | 48.0 | 48.3 | 48.5 |
| Alkynes | | | | |
| acetylene | 17.1 | 17.1 | 17.1 | 17.1 |
| propyne | 29.4 | 29.4 | 29.4 | 29.7 |
| Aromatics | | | | |
| benzene | 57.5 | 57.5 | 57.5 | 57.5 |
| toulene | 69.6 | 69.1 | 69.1 | 70.1 |
| naphtalene | 91.4 | 91.4 | 91.4 | 91.2 |

## 2.2. Details of molecular dynamics simulations
### 2.2.1 Details of simulations of electron impacts

The effect of electron irradiation on the GNR structure is simulated using the CompuTEM algorithm[21,22], which describes the structure evolution under electron irradiation as a sequence of irradiation-induced events (such as atom removal or bond rearrangement induced by an electron impact). Here we include into the CompuTEM algorithm the following steps: (1) the nanostructure is equilibrated at the temperature corresponding to the experimental conditions in HRTEM (an MD run at temperature 300 K for 10 ps); (2) a single electron–atom interaction event is introduced by giving a momentum distributed according to the standard theory of elastic electron scattering between a relativistic electron and a nucleus[33–35] to a random atom that is chosen based on the total cross-sections of electron collisions with different atoms (determined by the minimum transferred energies assigned to each atom), (3) an MD run is performed at temperature 300 K corresponding to the experimental conditions with the duration of 10 ps sufficient for bond reorganisation, (4) the surrounding of the impacted atom is analysed: if no change in the list of the nearest neighbours is detected (the impact is unsuccessful), the simulation cycle is repeated. But if the system topology has changed (the impact is successful), an additional MD run of duration 30 ps at the elevated temperature of 1500 K is performed to describe the structural relaxation between successive electron impacts.

The minimal transferred energies $T_{min}$ used for hydrogen and carbon atoms are 1 and 5 eV, respectively. The distributions of successful electron impacts producing structural changes over the energies transferred to carbon and hydrogen atoms show that nearly all the successful impacts occur at the transferred energies considerably greater than the chosen values of $T_{min}$ (see Figure S1 in Supplementary Data). Thus, almost all successful electron impacts are taken into account with such a choice of $T_{min}$. Neglecting the impacts with the energies below the minimal transferred ones (which virtually do not lead to any structural changes), the total cross-sections $\sigma_H$ and $\sigma_C$ of electron scattering for hydrogen and carbon atoms, respectively, can be calculated as functions of the kinetic energy $E_e$ of electrons in HRTEM. The differential scattering cross-section is computed according to the McKinley and Feshbach formula[35]:

$$\sigma(\theta) = \sigma_R \left[ 1 + \pi \frac{Ze^2}{\hbar c} - \beta^2 \sin^2\frac{\theta}{2} \beta \sin\frac{\theta}{2}\left(1 - \sin\frac{\theta}{2}\right) \right]$$
$$\sigma_R = \left(\frac{Ze^2}{2m_e c^2}\right)^2 \frac{1-\beta^2}{\beta^4} \csc^4\left(\frac{\theta}{2}\right)$$
(8)

where $Z$ is the nuclear charge, $\beta = V_e/c$ is the ratio of electron velocity $V_e$ to the speed of light $c$, $\hbar$ is the Planck constant, $e$ is the elementary charge, $\theta$ is the electron scattering angle, and $m_e$ is electron mass. The minimal angles for electron scattering corresponding to minimal transferred energies $T_{min}$ are evaluated as

$$\theta_{min} = 2\arcsin\left(\sqrt{\frac{T_{min}}{T_{max}}}\right) \quad (9)$$

and the total cross-section is calculated as

$$\sigma_{tot} = 2\pi \int_{\theta_{min}}^{\pi} \sigma(\theta)\sin\theta\, d\theta \qquad (10)$$

The dependences of the total cross-sections $\sigma_H$ and $\sigma_C$ for hydrogen and carbon atoms, respectively, on the electron energy $E_e$ obtained using Eq. (9) for the minimal transferred energies of 1 and 5 eV for hydrogen and carbon atoms, respectively, are shown in Figure 2.

The total structure evolution time under the experimental conditions in HRTEM can be expressed as a sum of the time periods between subsequent irradiation-induced events. The time period $t_{ev}$ between the events is defined as the inverse of the product of the overall cross-section $\sigma$ corresponding to all atoms of the system and the electron current density $j$ as $t_{ev} = 1/j\sigma$, where $\sigma = N_H\sigma_H + N_C\sigma_C$, $N_H$ and $N_C$ are the numbers of hydrogen and carbon atoms, respectively, in the system. This allows the direct comparison of the simulated and experimentally observed processes under electron irradiation (all impacts, including unsuccessful, which do not lead to bond rearrangements, are included in this expression). For all the electron energies considered in the present paper, the same value of the electron current density $j = 4.1 \times 10^6$ e$^-$/(nm$^2$s) is used. This electron current density lies within the range used for observation of formation of endohedral metallofullerenes from nickel clusters surrounding by amorphous carbon[15] and cutting of carbon nanotubes by nickel cluster[6] under electron irradiation in HRTEM. Below we show that such a choice of the electron current density should make possible visual observation of the structural transformation of the GNR into the hybrid 1D nanoobject.

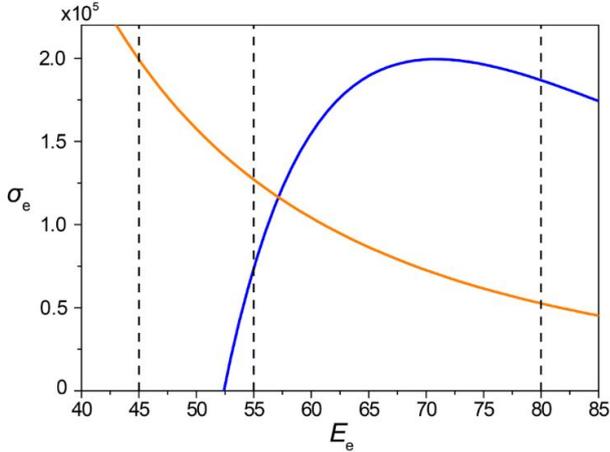

**Figure 2.** Calculated electron scattering cross-sections $\sigma_e$ (in barn) as functions of the kinetic energy $E_e$ (in keV) of electrons for hydrogen atoms and the minimum transferred energy of 1 eV (orange line), and carbon atoms and the minimum transferred energy of 5 eV (blue line). The kinetic energies of electrons used in the MD simulations (45, 55 and 80 keV) are shown using vertical dashed lines.

The kinetic energies of electrons considered in our MD simulations are chosen on the basis of Figure 2. The values 45, 55 and 80 keV correspond to different ratios of the total cross-sections for carbon and hydrogen atoms (computed using the minimal transferred energies listed above that allow to take into account virtually all successful impacts). Comparison of structure evolution under electron irradiation with these electron energies allows to distinguish the roles of spontaneous and irradiation-induced breaking and formation of bonds between carbon atoms during the formation of 1D nanoobjects. For the electron energy of 80 keV, the total cross-section for carbon atoms is considerably greater than for hydrogen atoms. Thus, a significant number of irradiation-induced events caused by electron impacts on carbon atoms should occur before hydrogen removal by electron impacts. For the electron energy of 55 keV, the total cross-section for hydrogen atoms is about twice greater than the one for carbon atoms. Therefore, the role of irradiation-induced events caused by electron impacts on carbon atoms in this case is considerably lower than for the electron energy of 80 keV. For the electron energy of 45 keV, electron impacts on carbon atoms are excluded as the total cross-section for them equals zero. This means that only spontaneous and thermally-induced breaking and formation of bonds between carbon atoms can occur under 45 keV electron irradiation after hydrogen removal.

### 2.2.2 Details of MD simulations

MD simulations have been carried out using the REBO-1990EVC_CH potential described in section 2.1. The in-house MD-kMC (Molecular Dynamics – kinetic Monte Carlo) code[36] is used for the simulations. The integration is performed following the velocity Verlet integration algorithm[37] with the time step of 0.6 fs. The temperature is maintained with the help of the Berendsen thermostat[38] with the relaxation time of 0.03 ps. Bond breaking and formation is detected through the analysis of the topology of the bond network using the "shortest-path" algorithm[39]. Two carbon atoms are considered as bonded if the distance between them lies within 1.9 Å, carbon and hydrogen atoms are considered as bonded if the distance between them lies within 1.4 Å.

### 2.3 Details of DFT calculations

To calculate energy characteristics of 1D nanoobjects obtained in our MD simulations we have performed spin-polarized DFT calculations using the VASP code[40] with the Perdew-Burke-Ernzerhof (PBE) exchange-correlation functional[41]. The interaction of valence electrons with atomic cores is described by the projector augmented-wave method[42]. The maximal kinetic energy of the plane-wave basis set is 400 eV. The second-order Methfessel-Paxton smearing[43] with the width of 0.1 eV is applied. The size of the simulation cell with periodic boundary conditions is $L \times 20 \times 20$ Å, where $L$ is the elementary period of the 1D nanoobject. For the chosen simulation cell, the distance between periodic images of the system in neighbour simulation cells exceeds 20 Å that is large enough to exclude possible interactions between them. The integration over the Brillouin zone is carried out using the Monkhorst-Pack method[44] with the 6×1×1 k-point grid. The structures are geometrically optimized until the residual force acting on each atom becomes less than 0.03 eV/Å. Barriers for transitions between different polycyclic regions are calculated using the nudged elastic band method (NEB)[45].

Spin-polarized calculations of electronic properties have been performed using Quantum Espresso[46—48]. The PBE functional is used. The first-order Methfessel-Paxton smearing[43] with the width of 0.07 eV is applied. The maximal kinetic energy of the plane-wave basis set is 816 eV. The kinetic energy cutoff for charge density and potential is 4500 eV. The stopping criterion for self-consistent cycles is $10^{-5}$ eV. The simulation cell of the same size as in the calculations of the barriers is considered. The 6×1×1 k-point grid is used for the ground state and the 60×1×1 k-point grid for band structure calculations. The positions of ions within the simulation cell are optimized till the maximum residual force becomes less than 0.05 eV/Å and the change in the total energy in successive iterations less than 3 meV.

### 3. Results

#### 3.1 Scheme of graphene nanoribbon synthesis

Let us consider a possible reaction scheme for the synthesis of a new type of GNRs with alternating regions of different widths displaying one and three hexagon rows, respectively. Among the most common reactions used in "on-surface synthesis"[20], the bottom-up approach whereby a great variety of atomically precise GNRs have been readily synthesized[18,19], we find Ullmann-coupling and dehydrogenation reactions. Herein we propose a new reactant that could use those two reactions to afford GNRs with periodically alternating sections of one and three hexagonal carbon rings across their width. The reactant consists in an appropriately functionalized octacene derivative (Figure 3a). Polyacenes featuring six or more rings display a partial open-shell character, which compromises their stability and makes their synthesis extremely challenging[49]. However, if some of the C atoms are functionalized in such a way that the conjugation is broken, the molecules gain back their stability and closed-shell character[49–51]. One of the approaches successfully used to break the conjugation is the double hydrogenation of the outer C atoms in a ring[49,51,52], as proposed in Figure 3a on the third ring from either side. On the other hand, the halogen atoms on the two terminal rings on either side of the molecule allow for the polycondensation via Ullmann coupling[20], which would result in the polymer shown in Figure 3b. Further annealing of the polymer would eventually trigger the dehydrogenation of the $sp^3$ carbon atoms[51,52] and the ultimate formation of the unsubstituted and fully conjugated GNR displayed in Figure 3c.

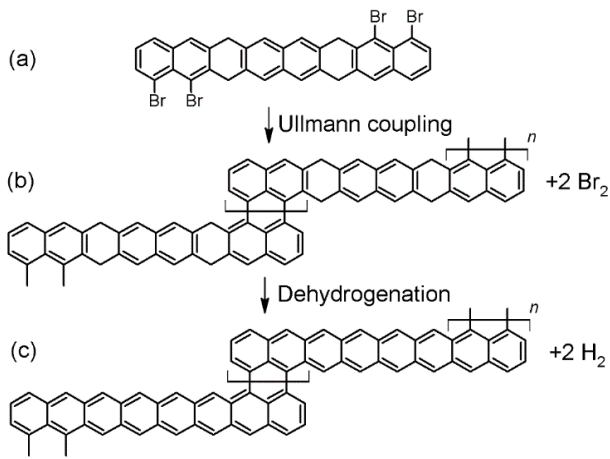

**Figure 3.** (a) Proposed reactant molecule, a hydrogenated and brominated octacene derivative. (b) Polymeric structure obtained from the Ullmann coupling of the reactant. (c) Fully conjugated GNR structure obtained after dehydrogenation of the polymer.

#### 3.2 MD simulations of hybrid 1D nanoobjects formation

To study the formation of hybrid 1D nanoobjects with alternating double chains and polycyclic graphene-like regions under electron irradiation of the GNRs in HRTEM we have performed reactive MD simulations using the REBO-1990EVC_CH potential and CompuTEM algorithm described above. In total, we have performed 10 MD simulation runs for each of the considered electron kinetic energies (80, 55 and 45 keV) starting from the chosen initial GNR based on the structure described in the previous chapter and consisting of 7 elementary periods. One elementary period includes one narrow and one wide region of the proposed GNR, see the upper structure shown in Figure 4a. The length of the elementary period is 16.3 Å and the total length of the GNR is 125.6 Å. Twelve atoms at each end of the GNR were kept fixed during the simulations. All simulation runs for the electron energy of 80 keV were performed till the rupture of the forming 1D nanoobject. All simulation runs for the electron energy of 55 keV were carried out for 500 s, which is 10 times longer than the minimum time of hydrogen removal from a single elementary period of the GNR. Indeed all hydrogen atoms were removed by electron impacts from the forming 1D nanoobject within this time. All simulation runs for electron energy 45 keV were performed till removal of all hydrogen atoms from the forming 1D nanoobject since electron impacts on carbon atoms do not lead to any structural changes at this electron energy.

The visual analysis of structure evolution reveals formation of alternating double atomic carbon chains and polycyclic regions from narrow and wide regions of the initial GNR, respectively, after removal of hydrogen atoms by electron impacts for all the considered kinetic energies of electrons of 45, 55 and 80 keV. The typical examples of structure evolution of the GNR with edges terminated by hydrogen atoms under 80 and 55 keV electron irradiation in HRTEM observed in the MD simulations are shown in Figures 4a and 5a, respectively. Transformation of the GNR structure under electron irradiation for the electron energy of 45 keV occurs in the similar manner. According to the DFT calculations,[10] 1-ZGNR is unstable and should decompose into double atomic carbon chains. This explains why double atomic carbon chains form after hydrogen removal for all the considered kinetic energies of electrons. However, the structure of polycyclic regions formed under irradiation by electrons with energies 45 and 55 keV considerably differs from their structure formed under 80 keV electron irradiation. In particular, the polycyclic regions formed under 45 and 55 keV electron irradiation contain on average more polygons and the diversity of such regions is reduced in comparison with polycyclic regions formed under 80 keV electron irradiation. Figures 4b and 5b show that number of successful electron impacts on carbon atoms, i.e. those producing structural changes, before removal of all hydrogen atoms is considerably greater for the case of electron irradiation with the largest considered energy of 80 keV. Note that for the electron kinetic energies of 55 and 80 keV, the ratio of the number of removed hydrogen atoms to the number of successful electron impacts on carbon atoms correlates excellently with the calculated ratio of electron scattering cross-sections for hydrogen and carbon atoms shown in Figure 2. The successful electron impacts on carbon atoms are completely absent under 45 keV electron irradiation. It is clear that successful electron impacts on carbon atoms prevent formation of the polycyclic regions which contain from 5 to 7 polygons. Moreover, successful electron impacts on carbon atoms eventually lead to rupture of the forming 1D nanoobject. In the case of the GNR transformation under 80 keV electron irradiation, the average rupture time is 164 ± 12 s. About 80 % of the 1D nanoobject parts corresponding to the former elementary periods of the initial GNR still contain at least one hydrogen atom at the rupture moment.

Let us consider the atomistic mechanism of formation of 1D nanoobjects and statistics for different structures of polycyclic regions obtained on the basis of simulation runs for transformation of the GNR structure under 45 and 55 keV electron irradiation. The average time of total hydrogen removal from one elementary period of the GNR is 119 ± 8 and 72 ± 3 s for the kinetic energies of electrons of 55 and 45 keV, respectively. The ratio of these times equals 1.65, which is in excellent agreement with the ratio of calculated electron scattering cross-sections for hydrogen at corresponding electron kinetic energies (which is 1.56, see Figure 2). It should be mentioned, however, that the time of total hydrogen removal considerably differs for elementary periods of the GNR and, for example, ranges from 24 to 439 s for the kinetic energy of electrons of 55 keV. Formation of double atomic carbon chains and polycyclic regions takes place after hydrogen removal from the narrow and wide regions of the initial GNR, respectively. Note that formation of double chains can occur before the total hydrogen removal from the wide regions of the GNR.

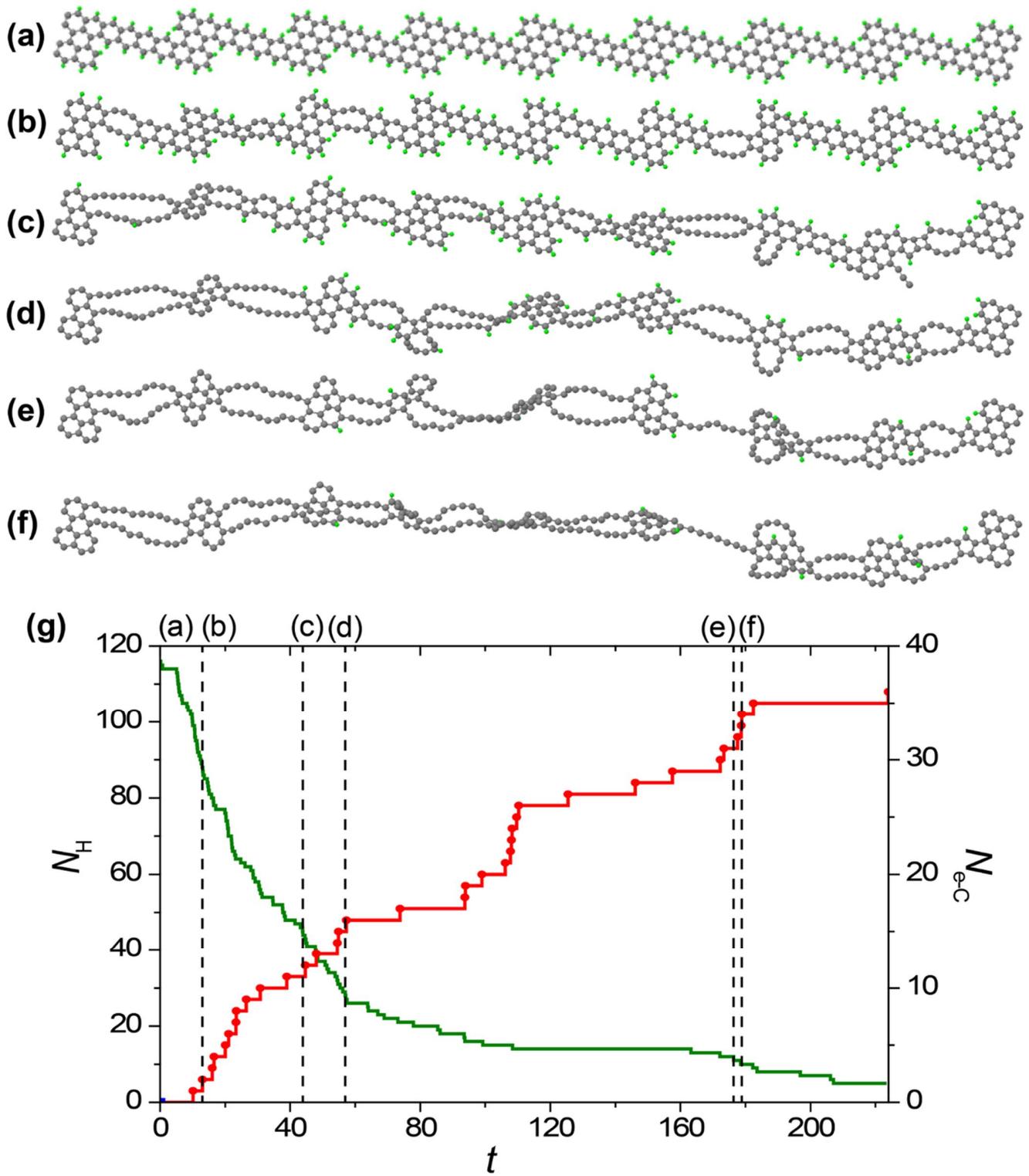

**Figure 4.** (a-f) Simulated structure evolution of the graphene nanoribbon with edges terminated by hydrogen atoms (shown in green) under 80 keV electron irradiation in HRTEM: (a) 0 s, (b) 13.0 s, (c) 43.9 s, (d) 56.9 s, (e) 176.3 s and (f) 178.9. Nanoribbon rupture occurs at 223.9 s. (g) Calculated number $N_H$ of hydrogen atoms remaining attached to the structure (green line) and number $N_{e\text{-}C}$ of successful electron impacts on carbon atoms (red line) as functions of time $t$ in the same simulation run. The moments of time corresponding to structures (a–f) are shown using vertical dashed lines.

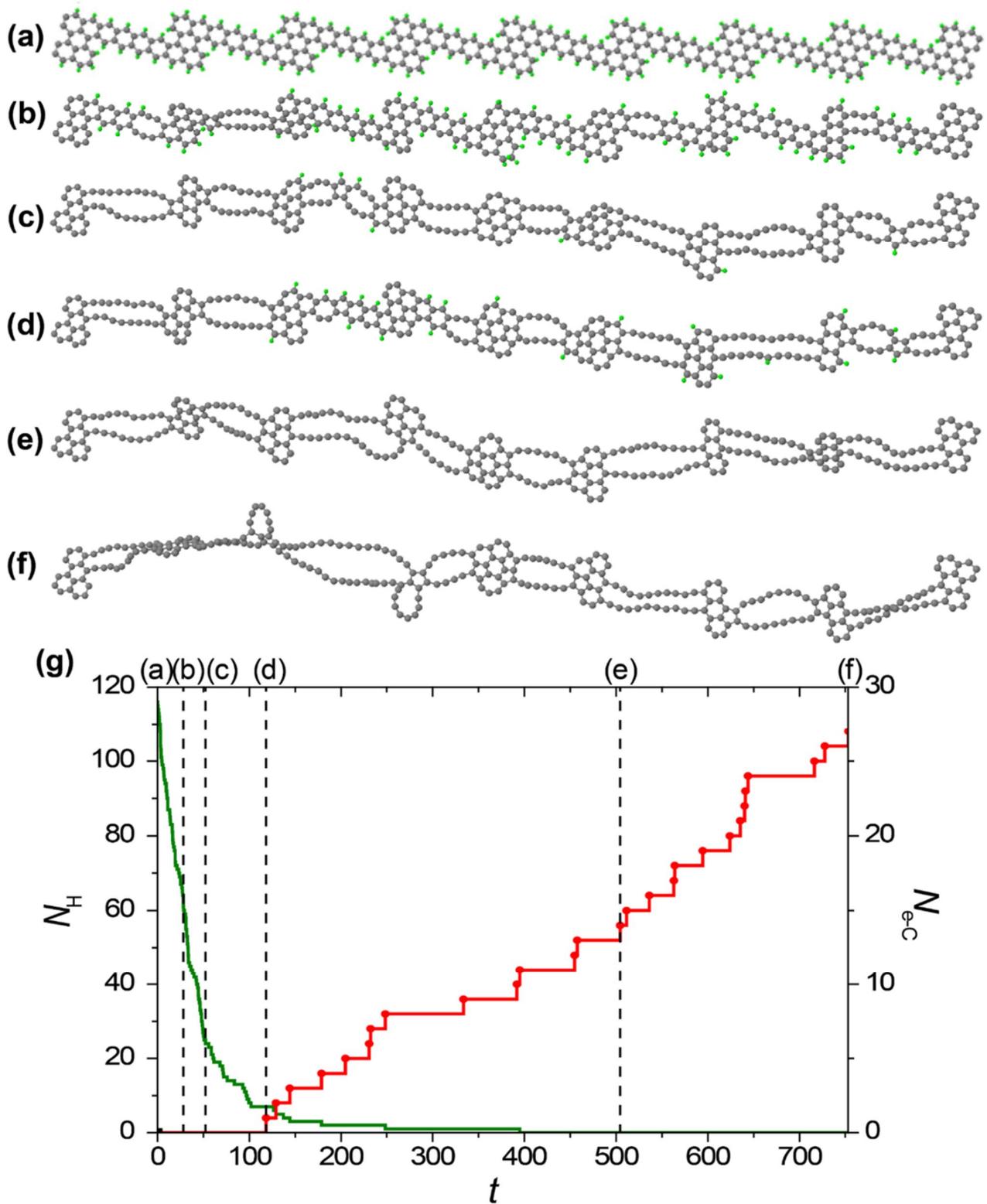

**Figure 5**. (a-f) Simulated structure evolution of the graphene nanoribbon with edges terminated by hydrogen atoms (shown in green) under 55 keV electron irradiation in HRTEM observed in: (a) 0 s, (b) 27.9 s, (c) 52.3 s, (d) 118.2 s, (e) 504.5 s and (f) 753.4 s. (g) Calculated number $N_H$ of hydrogen atoms remaining attached to the structure (green line) and number $N_{e-C}$ of successful electron impacts on carbon atoms (red line) as function of time $t$ in the same simulation run. The moments of time corresponding to structures (a–f) are shown using vertical dashed lines.

After hydrogen removal from the wide regions of the GNR, formation of new bonds between carbon atoms leading to an increase in the number of polygons in polycyclic regions of the forming 1D nanoobject from 3 hexagons (in the initial structure of the wide region of the GNR) to the limit of 7 polygons is observed. Subsequent stages of polycyclic region growth and notations assigned to the polycyclic regions of different structure along the growth path are shown in Figure 6a. All 4 new polygons formed as a result of bond formation after hydrogen removal are pentagons. Not only bond formation but also breaking of these bonds both due to electron impacts and thermal activation have been observed for all the reactions shown in Figure 6a. To determine abundant polycyclic regions, the fractions of these regions have been calculated taking into account only the number and relative positions of the polygons in accordance with Figure 6a (that is the calculated fractions include several structures with different numbers of atoms in the polygons, which are formed occasionally due to electron impacts on carbon atoms under 55 and 80 keV electron irradiation). Figure 6b shows these calculated fractions of different polycyclic graphene-like regions as functions of time during 1D nanoobject formation for all the considered kinetic energies of electrons of 45, 55 and 80 keV. Evidently, the low fraction of the polycyclic regions corresponding to the revealed growth path under 80 keV electron irradiation is related to the considerable number of electron impacts on carbon atoms leading to structural changes which spoil the structure of polycyclic regions (see Figure 6c). In spite of the presence of some electron impacts on carbon atoms under 55 keV electron irradiation, the calculated fractions of 1D nanoobjects with polycyclic regions D1, D2, E and F are the same for the kinetic energies of electrons of 45 and 55 keV at the same moments of time within statistical errors. The 1D nanoobject with the polycyclic regions D1 which consists of 5 polygons is found to be the most abundant for these electron energies with the fraction 25–30 % at the moment of about 200 s, when hydrogen is almost removed. The total amount of four abundant polycyclic regions shown in Figure 6a, D1 and D2 (5 polygons), E (6 polygons) and F (7 polygons), is up to 80 % of all polycyclic regions formed under 45 keV electron irradiation. The analysis of transitions observed between the 1D nanoobjects with the polycyclic regions corresponding to the growth path shows that the greater fraction of the 1D nanoobjects with the polycyclic regions C formed under 55 keV electron irradiation in comparison with this fraction under 45 keV electron irradiation is explained by the transitions from the 1D nanoobjects with the polycyclic regions D1 induced by electron impacts on carbon atoms.

### 3.3 DFT study of hybrid 1D nanoobjects

The detailed DFT study of structure and electronic properties of the hybrid 1D nanoobjects with different polycyclic regions revealed here in the MD simulations will be performed elsewhere. Here we restrict ourselves to consideration of energetics of the polycyclic regions observed and barriers for reactions of bond formation along the revealed growth path presented in Figure 6a. Additionally electronic properties of the 1D nanoobject with polycyclic regions of the most abundant

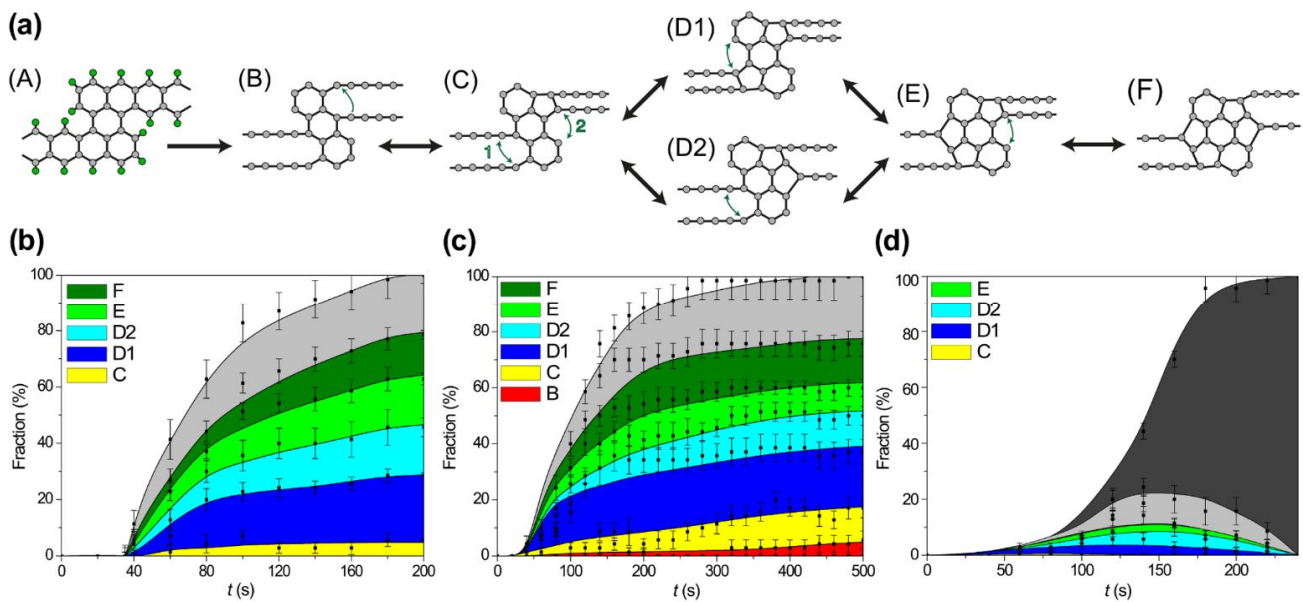

**Figure 6.** (a) Scheme of formation of a one-dimensional nanoobject with alternating double chains and polycyclic graphene-like regions from a graphene nanoribbon under electron irradiation in HRTEM. (A) Initial structure of the graphene nanoribbon, (B) structure after removal of hydrogen atoms, (C) – (F) abundant structures observed after subsequent bond formation for the growth path of polycyclic graphene-like regions. Green arrows point to atoms that form a new bond. Reactions where both formation and breaking of bonds have been observed are indicated by double-headed arrows. (b) – (d) Calculated fractions of different structures of polycyclic graphene-like regions as functions of time t under (a) 45, (b) 55 and (c) 80 keV electron irradiation in HRTEM: any structure before removal of all hydrogen atoms (white), structures after removal of hydrogen atoms with the number and relative positions of polygons shown in scheme (a): B (red), C (yellow), D1 (dark blue), D2 (light blue), E (light green), F (dark green), other structures (light grey) and structures of a broken ribbon (dark grey).

structure D1 are studied.

### 3.3.1. Energetics of the polycyclic region growth path

To find the ground state of 1D nanoobjects with polycyclic regions B, D1 and F, several calculations have been performed for each type of the polycyclic region with the size $L$ of the simulation cell varied with a step of $\Delta L = 0.05$ Å starting from $L = 26.42$ Å. The calculated ground-state energies relative to the ground-state energy of the 1D nanoobject with polycyclic regions B, which correspond to the beginning of polycyclic region growth path, and the optimal elementary periods $L_0$ corresponding to the ground states of the 1D nanoobjects with different polycyclic regions are given in Table 4. We have found that formation of new bonds during the polycyclic region growth leads to the decrease of the ground-state energy of the 1D nanoobject and the nanoobject with the polycyclic regions F composed of 7 polygons is the most favorable. The smaller optimal elementary period of the 1D nanoobject with polycyclic regions F in comparison with the periods of the 1D nanoobjects with polycyclic regions B and D1 is caused by the curved shape of the polycyclic region F. Such a curved shape of the polycyclic region F is analogous to the curved shape of the polycyclic aromatic hydrocarbon corannulene with a pentagon in the structure[53].

**Table 4.** Ground-state energies $E_0$ relative to the 1D nanoobject with polycyclic regions B and optimal elementary periods $L_0$ obtained by the DFT calculations for the 1D nanoobjects with the polycyclic regions B, D1 and F.

| Polycyclic region | $L_0$, Å | $E_0$, eV |
| --- | --- | --- |
| B | 26.32±0.05 | 0 |
| D1 | 26.32±0.05 | –2.74 |
| F | 24.32±0.05 | –4.53 |

To confirm the atomistic mechanism of 1D-nanoobject formation observed in the MD simulations, the energetics of each step of the polycyclic region growth path shown in Figure 6a has been studied by the NEB method via both DFT and empirical potential-based calculations. Because of the restrictions of the NEB method, the same length of the simulation cell 26.32 Å along the 1D-nanoobject axis has been used in all calculations. This length of the simulation cell corresponds to the calculated optimal elementary periods of the 1D nanoobjects with the polycyclic regions B and D1. The calculations with the fixed length of the simulation cell can correspond to the experimental conditions where the initial GNR is suspended with the ends fixed (see, for example, Ref. 54 on the GNR-based transistor). The calculated energies $E_1$ of the 1D nanoobjects with the polycyclic regions corresponding to the growth path and related barriers for reactions of bond formation are given in Table 5 and Table 6, respectively. Figure 7 shows a schematic representation of energetics along the 1D nanoobject growth path that follows from the DFT calculations.

According to the DFT calculations, formation of new bonds between carbon atoms during the polycyclic region growth leads to the decrease of the 1D nanoobject energy by about 0.5-1 eV per one new bond. Note that the energy of the 1D nanoobject with the polycyclic regions F and the elementary period 26.32 Å is 0.95 eV greater than the energy of this nanoobject with the optimal elementary period 24.32 Å. Both the DFT calculations and those using the empirical potential show that the flat 1D nanoobject with the polycyclic regions C is unstable. This explains the small fraction of the 1D nanoobject with the polycyclic regions C observed in our MD simulations of the GNR transformation under 45 keV electron irradiation, where successful electron impacts on carbon atoms leading to structural changes are absent. The observation of the 1D nanoobject with the polycyclic regions C in our MD simulations can be possibly attributed to metastability of this nanoobject if it is twisted or bended at nonzero temperature (see Figure 5f and video in Supplementary Data). All the barriers for new bond formation along the 1D-nanoobject growth path obtained by the DFT calculations are within 0.55 eV. These small values of the barriers and considerable decrease of the 1D-nanoobject energy along the growth path excellently confirm the atomistic mechanism of the 1D nanoobject formation revealed by the MD simulations using the classical empirical potential.

**Table 5.** Energies $E_1$ (in eV) of the 1D nanoobjects with different polycyclic regions relative to the 1D nanoobject with polycyclic regions B obtained by the DFT calculations and using the empirical potential REBO-1990EVC_CH.

| Polycyclic region | $E_1$, eV | |
| --- | --- | --- |
| | DFT (PAW PBE) | Potential REBO-1990EVC_CH |
| B | 0 | 0 |
| C | unstable | unstable |
| D1 | –2.74 | –3.75 |
| D2 | –2.29 | –3.70 |
| E | –3.13 | –5.41 |
| F | –3.58 | –7.84 |

**Table 6.** Barriers $E_a$ for the transitions between the 1D nanoobjects with the polycyclic regions B–F (in eV) obtained by the DFT calculations and using empirical potential REBO-1990EVC_CH.

| Reaction | $E_a$, eV | |
| --- | --- | --- |
| | DFT (PAW PBE) | REBO-1990EVC_CH |
| B→D1 | 0 | 0.22 |
| B→D2 | 0.35 | 0.21 |
| D1→E | 0.44 | 0.12 |
| D2→E | 1.92 | 2.03 |
| E→F | 0.55 | 0.14 |

Although the calculations based on the empirical potential overestimate the energy decrease along the 1D-nanoobject growth path, the differences between the values of the barriers obtained by DFT calculations and using the potential are with-in 0.4 eV. It is a good accuracy for reaction barrier estimates using a reactive empirical potential. Thus, the REBO-1990EVC_CH potential developed here is adequate for atomistic simulations of GNR transformations.

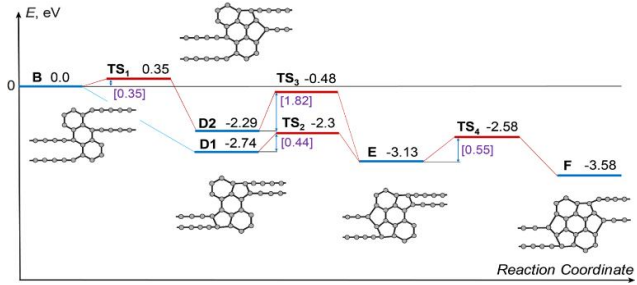

**Figure 7**. Schematic representation of energetics along the revealed growth path presented in Figure 6a following from the DFT calculations. All the calculated energies, including the energies of transition states TS, are given relative to the energy of the 1D nanoobject with the polycyclic regions of type B, which is set to zero. The values of the barriers of transitions between the nanoobjects with polycyclic regions of different types are indicated in square brackets.

### 3.3.2. Electronic properties of the hybrid 1D nanoobject with D1 polycyclic regions

Electronic properties of GNRs[30,55-58] and atomic carbon chains[59–69] have been recently in the focus of numerous studies aimed at development of carbon-based nanoelectronics (see, e.g.,[70] for review). To get insight into the electronic properties of the new hybrid 1D nanoobject proposed in the present paper, we have performed the DFT calculations for the system with polycyclic regions of the most abundant structure D1 consisting of 5 polygons (Figure 7 and Figure S2 of Supplementary Data). Our DFT calculations show that the band structure of the nanoobject (Figure 8) is determined by the magnetic ordering at edges of the polycyclic regions, similar to zigzag GNRs[30,55–57]. In the non-magnetic (NM) state (Figure 8c), we observe band crossing at the Fermi level and nearly flat bands close to the Fermi level originating from edge states of the polycyclic regions. These flat bands provide a sharp peak in the density of states at the Fermi level indicating the electronic instability of the system. Correspondingly, we find that the NM state is unstable against antiferromagnetic (AFM) and ferromagnetic (FM) ordering with antiparallel and parallel spin orientations at the edges of the polycyclic regions, respectively. In the AFM state (Figure 8b), a band gap is opened. In the FM state (Figure 8a), there are bands crossing the Fermi level. However, the flat bands are moved away from the Fermi level.

According to our calculations, the ground state of the 1D nanoobject corresponds to the FM state. However, the energy difference between the AFM and FM states is only 0.9 meV per elementary period of the 1D nanoobject. The NM state is much more unstable, with the relative energy with respect to the FM state of 37 meV per elementary period. The 1D nanoobject in the AFM state has an indirect band gap (Figure 8b) with the maximum of the valence band at the $\Gamma$ point $(0,0,0)$ and minimum of the conductance band at the X point $(\pi/L_0, 0, 0)$. This band gap is very small, only 0.06 eV. The direct band gaps along the $\Gamma - X$ path lie in the range from 0.17 eV to 0.25 eV.

The analysis of partial densities of states for atoms of the 1D nanoobject reveals that the bands close to the Fermi level originate mostly from $p_z$ orbitals (Figure 9 for the FM state and Figure S3 of Supplementary Data for the AFM state; the $z$ axis is directed perpendicular to the ribbon plane). The shapes of partial densities of states for neighbor atoms alternate both in the chains and in the polycyclic regions. It is confirmed that the flat bands at 0.2–0.3 eV relative to the Fermi level in the FM and AFM states are mostly formed by the $s$, $p_x$ and $p_y$ orbitals of two-coordinated carbon atoms at the edges of the polycyclic regions, more exactly from the first and third ones in the three-atom chains at the edges (see the results for atom #31 in Figure 9 and Figure S3 of Supplementary Data; the $x$ axis is directed along the ribbon axis and the y axis is perpendicular to the $x$ and $z$ axes). There is also a significant contribution from the three-coordinated atom that is the second-order neighbor of these atoms within the same hexagonal ring (atom #28). The flat bands at about −0.6 eV relative to the Fermi level in the FM and AFM states are formed mostly by $p_x$ and $p_y$ orbitals of atoms in the chains (e.g., atoms #3 and #5). The lower-lying flat bands at about −0.8 eV come from $p_z$ orbitals of atoms both in the chains and within the polycyclic regions.

The effect of edge magnetization on the band structure is well studied for zigzag GNRs[30,55-58]. The presence of the magnetic order at edges of zigzag GNRs was demonstrated experimentally at room temperature for ribbons up to 7 nm in width[58]. According to the DFT calculations[55], the energy difference between the FM and AFM states for the widest of such ribbons can be as small as 0.4 meV per unit cell. This suggests that it might be possible to observe the magnetic effects in the 1D nanoobject studied here. Note that for zigzag GNRs[30,55-57], the AFM state is preferred over the FM one, different from the 1D nanoobject. Band gaps of several hundreds of meV have been reported for narrow zigzag GNRs with pristine edges in the AFM state[30,55,56.] They are much greater than the indirect band gap found for the 1D nanoobject but comparable to the direct band gaps.

Our calculations show that the bond lengths in the chains of the 1D nanoobject composed of 16 atoms each alternate because of the Peierls distortion[11,59–64] (Figure S2 of Supplementary Data). Correspondingly, there is a variation in the bond order along the chains, which can be appreciated from the distribution of the electron density (Figure 10). Thus, the chains in the 1D nanoobject are polyyne-like. The bond length alternations of computed according to papers[59,60] are found to be 6.0 pm and 5.8 pm for the two chains within the elementary period of the nanoobject (eq. (S1) of Supplementary Data). Such a bond length alternation gives rise to a band gap in isolated carbon chains[11,59–64]. The 1D nanoobject studied here, however, is metallic in the FM state, while the band gap in the AFM state is extremely small. Therefore, the chains connected the polycyclic region within the 1D nanoobject display a metallic behavior. A significant metal-like conductivity has been reported previously for polyyne-like chains connecting metal electrodes, where a charge transfer from the metal to the chains takes place[65,66]. A charge transfer should also occur in the 1D nanoobject between the chains and polycyclic regions. Note that the structure of the polycyclic regions is destabilized by the presence of the dangling bonds at the edges. Stabilization of the edges of the polycyclic regions through passivation with hydrogen should affect the charge transfer and might change the character of the 1D nanoobject to semiconducting.

The spatial distributions of the charge and magnetization densities over the 1D nanoobject in the FM and AFM states are shown in Figure 10. It is seen that the magnetization density is mostly localized at edge atoms of the polycyclic regions. The signs of the magnetic moments alternate for neighbor atoms within the polycyclic regions, the same as for zigzag GNRs[55]. The magnetic moments of the atoms in the chains are very small, in agreement with previous findings for polyyne-like carbon chains with an even number of atoms[59–61] (note that the situation is different in odd chains, where the bond alternation cannot be fulfilled everywhere: uncompensated delocalized charge and, correspondingly, a nonzero magnetic moment arise there from switching of bond type from single/triple at the chain ends to double/double in the middle of the chain[60]). The total magnetization of the 1D nanoobject in the FM state is found to be $2.8\mu_B$, where $\mu_B$ is the Bohr magneton.

Spin transport through atomic carbon chains connecting GNR electrodes has been recently studied in connection with the potential of such systems for spintronic devices[60,67,68]. We observe that for the ground-state FM ordering, the band structures of the 1D nanoobject are rather different for spin up and spin down (Figure 8a). It is seen from these band structures that it might be possible to shift the Fermi level by about 0.1 eV, e.g., by doping or application of a gate potential, to the energy where the density of states for one of the spin channels is zero and for the other is not. In this case, the 1D nanoobject can be metallic for one of the spin orientations and semiconducting for the other, which makes this structure interesting for spintronic applications.

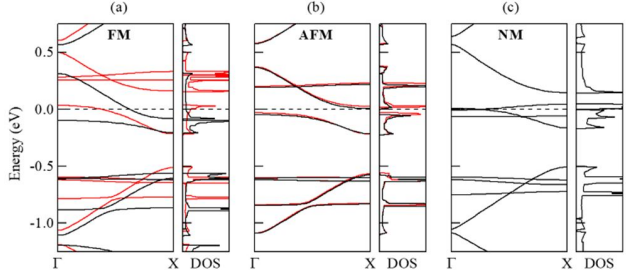

**Figure 8.** Computed band structures and densities of states (DOS) of the 1D nanoobject with D1 polycyclic regions in the (a) ferromagnetic, (b) antiferromagnetic, and (c) nonmagnetic states. The energies in eV are given relative to the Fermi level. The results for spin up and down are shown by black and red lines, respectively.

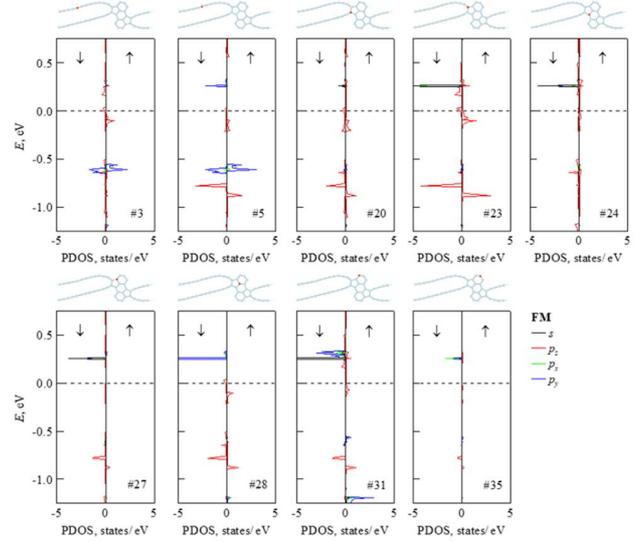

**Figure 9.** Projected densities of states for selected atoms of the 1D nanoobject with D1 polycyclic regions (as indicated in the atomistic structures) in the ferromagnetic state. The energies in eV are given relative to the Fermi level. The results for $s$, $p_z$, $p_x$ and $p_y$ orbitals are shown by black, red, green and blue lines, respectively.

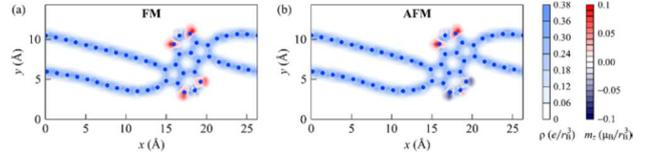

**Figure 10.** Calculated spatial distributions of the charge density (in $e/r_B^3$, where $e$ is the elementary charge and $r_B^3$ is the Bohr radius; light blue) and magnetization density (in $\mu_B/r_B^3$, where $\mu_B$ is the Bohr magneton; red and dark blue) of the 1D nanoobject with D1 polycyclic regions in the (a) ferromagnetic and (b) antiferromagnetic states. The coordinates are given in Å.

**Conclusions**

The present paper is devoted to the reactive MD study of formation of the new hybrid type of 1D nanoobjects consisting of alternating double atomic carbon chains and graphene-like polycyclic regions from initial GNRs of alternating width. The scheme of synthesis of the GNR with alternating regions that are single and three hexagons wide using Ullmann coupling and dehydrogenation reactions is proposed. It is worth keeping in mind that the proposed GNR structure could be easily varied by modifying the starting reactant. Although we have herein proposed the use of an octacene derivative, longer or shorter polyacene derivatives could be used to increase or decrease the length of the narrow single hexagon wide region. Indeed, following similar on-surface synthesis concepts as proposed here, unsubstituted polyacenes with as many as twelve rings have been successfully synthesized[71], which could equally well be used as the repeating unit in the GNR formation. Assuming a similar behaviour upon irradiation with high energy electrons, the length of the single atoms chains linking the polycyclic regions could thus be controllably varied.

We have improved the reactive Brenner potential[24] for carbon-hydrogen systems. In our previous paper[16,25], we have reparameterized this potential to accurately describe energies of pristine graphene edges, vacancy migration in graphene and formation of atomic carbon chains. Here we have also extended the potential to the case of graphene edges functionalized by hydrogen, which is important for adequate modeling of the system studied.

The MD simulations performed with this advanced potential reveal formation of double chains and polycyclic regions from the narrow and wide regions of the proposed initial GNR, respectively, under electron irradiation in HRTEM for all the considered electron energies of 45, 55 and 80 keV. Thus, the 1D nanoobject with alternating double chains and polycyclic regions is formed from the initial GNR. The atomistic mechanism of the 1D nanoobject formation is found to be the following. Firstly hydrogen is removed by electron impacts. After that breaking and formation of bonds between carbon atoms occur for narrow and wide regions of the GNR, respectively. Namely, the narrow regions of the initial GNR spontaneously decompose into the double chains (the zigzag GNR with only one hexagon row in width is unstable[10]). Simultaneously the wide regions of the initial GNR grow from initial 3 hexagons to polycyclic regions composed of 5 to 7 polygons with 5 different polycyclic regions found for path of this growth. It is of interest that all the newly formed polygons are pentagons.

The fractions of polycyclic regions corresponding to the growth path of the 1D nanoobject formed are calculated. The maximal yield of abundant polycyclic regions is achieved at the lowest electron energies of 45 and 55 keV. Namely the total yield of 4 abundant polycyclic regions is about 80 %, while the yield of the most abundant polycyclic region which consists of 5 polygons reaches up to 30 %. This is a rather high yield in comparison with the typical yield in synthesis of various nanoobjects. For example, the yield of 13 % has been achieved for the synthesis of carbon rings from cyclocarbon oxide $C_{24}O_6$[72]. Reactions induced by electron impacts on carbon atoms are absent under 45 keV electron irradiation. This means that formation of new bonds between carbon atoms of polycyclic regions occurs spontaneously or due to thermal activation. On the contrary, the total yield of abundant polycyclic regions is very low (less than 10%) under high 80 keV electron irradiation since electron impacts on carbon atoms lead to etching of polycyclic regions and eventually to rupture of the forming 1D nanoobject. Note that formation of graphene-like polycyclic regions efficiently occurs without electron impacts on carbon atoms in contrast with reconstruction of zigzag graphene edges[2] or narrow GNRs[1] under electron irradiation. Thus, to produce 1D nanoobjects, the electron energy should correspond to the range where hydrogen atoms are efficiently removed by electron impacts but electron impacts on carbon atoms are almost absent and the choice of the electron energy in HRTEM cannot be used to increase the yield of abundant polycyclic regions.

To confirm the atomistic mechanism of formation of the 1D nanoobject revealed by the MD simulations, energy differences and activation barriers for the transitions with new bond formation along the growth path of polycyclic regions have been obtained using the DFT calculations. All the barriers for the transitions along the polycyclic region growth path observed in the MD simulations are within 0.55 eV. It is found that some bonds are formed without a barrier. This confirms that formation of the 1D nanoobject with the polycyclic regions can easily take place after hydrogen removal by electron impacts. Note that the differences in the values of the barriers obtained by the DFT calculations and using the advanced reactive potential are within 0.4 eV.

The DFT calculations performed show that formation of each new bond along the polycyclic region growth path leads to a decrease of the 1D nanoobject energy. That is the lowest energy corresponds to the 1D nanoobject with the polycyclic region consisting of 7 polygons. Therefore, the yield of the 1D nanoobject with this polycyclic region can be increased by control of the temperature regime during or after electron irradiation in HRTEM. Since strain affects the energy difference between different polycyclic regions and activation barriers for transitions between them, it can be also used to increase the yield of given polycyclic regions. Note that application of strain[12] or heating[11] simultaneously with the electron irradiation influences the yield of atomic carbon chains formed from GNRs.

The DFT calculations of the 1D nanoobject with the polycyclic regions of the most abundant structure composed of 5 polygons show that the non-magnetic state is unstable against magnetic ordering of spins at two edges of the polycyclic regions similar to the phenomenon observed in zigzag GNRs[30,55–58]. In the ground state, the 1D nanoobject is found to be ferromagnetic with parallel spins at the edges of the polycyclic regions. However, the aniferromagnetic state with antiparallel spins at the edges of the polycyclic regions is only slightly higher in energy. The chains of the 1D nanoobject composed of an even number of atoms display virtually no magnetism. The calculations of the band structures suggest that in the ferromagnetic state, the 1D nanoobject should behave as a metal, while in the antiferromagnetic state, a small band gap is opened. The band structures computed for spins up and down in the ferromagnetic case are rather different. There are energy windows where the density of states is nonzero for only one spin channel. This opens possibilities for the use of the 1D nanoobject in spintronic applications.

We believe that the synthesis of hybrid 1D nanoobjects proposed in the present paper can be considered as an example of the following general three-stage strategy of synthesis of new nanoobjects or macromolecules. Firstly precursors of organic nanoobjects or macromolecules are synthesized by the traditional chemical methods. Then these precursors are placed in HRTEM with the electron energy that is sufficient to remove hydrogen atoms but is not sufficient to cause breaking or forming of bonds between atoms heavier than hydrogen. As a result of hydrogen removal, the precursors become unstable or metastable and transform spontaneously or due to thermal activation into new nanoobjects or macromolecules which cannot be produced by the traditional chemical methods. Previously the analogous process of formation of crosslinking bonds between polymer molecules after hydrogen removal by electron impacts has been observed only for 3D samples of polymers[73,74]. Here we propose such a way of synthesis for production of isolated nanoobjects and macromolecules.

## ASSOCIATED CONTENT

Electronic Supplementary Information (ESI) available: figures illustrating distributions of successful electron impacts over the energies transferred to hydrogen and carbon atoms for the chosen minimal transferred energies, figures illustrating DFT studies of the 1D nanoobject with D1 polycyclic regions: bond length alternation in carbon chains and partial densities of states in the antiferromagnetic state, video files showing examples of the structure evolution observed in molecular dynamics simulations of the 1D nanoobject formation from the graphene nanoribbon with alternating width.

## AUTHOR INFORMATION

**Corresponding Author**

Andrey M. Popov − Institute for Spectroscopy of Russian Academy of Sciences, Troitsk, Moscow 108840, Russia; Email: popov-isan@mail.ru

## ACKNOWLEDGMENT

ASS, AMP and AAK acknowledge the Russian Foundation of Basic Research (Grants 18-02-00985 and 20-52-00035). SVR and NAP acknowledge the Belarusian Republican Foundation for Fundamental Research (Grant No. F20R-301) and Belarusian National Research Program "Convergence-2020". This work has been carried out using computing resources of the federal collective usage center Complex for Simulation and Data Processing for Mega-science Facilities at NRC "Kurchatov Institute", http://ckp.nrcki.ru/ and was supported by the Research Center "Kurchatov Institute (order No. 1569 of July 16, 2019). DGO has received funding from the Spanish Agencia Estatal de Investigación (Grant Nos. PID2019-107338RB-C63).

# Electronic supplementary information for

# Transformation of a graphene nanoribbon into a hybrid 1D nanoobject with alternating double chains and polycyclic regions

Alexander S. Sinitsa[a], Irina V. Lebedeva[b], Yulia G. Polynskaya[c], Dimas G. de Oteyza[d,e,f], Sergey V. Ratkevich[g], Andrey A. Knizhnik[c], Andrey M. Popov[i,1], Nikolai A. Poklonski[g] and Yurii E. Lozovik[i,j]

[a] National Research Centre "Kurchatov Institute", Kurchatov Square 1, Moscow 123182, Russia.

[b] CIC nanoGUNE BRTA, Avenida de Tolosa 76, San Sebastian 20018, Spain.

[c] Kintech Lab Ltd., 3rd Khoroshevskaya Street 12, Moscow 123298, Russia.

[d] Donostia International Physics Center, San Sebastián, Spain

[e] Centro de Física de Materiales (CFM-MPC), CSIC-UPV/EHU, San Sebastián, Spain

[f] Ikerbasque, Basque Foundation for Science, Bilbao, Spain.

[g] Physics Department, Belarusian State University, Nezavisimosti Ave. 4, Minsk, 220030, Belarus

[i] Institute for Spectroscopy of Russian Academy of Sciences, Fizicheskaya Street 5, Troitsk, Moscow 108840, Russia.

[j] National Research University Higher School of Economics, 109028 Moscow, Russia

**Contents**
**p. S2:** Distributions of successful electron impacts over the energies transferred to hydrogen and carbon atoms for the chosen minimal transferred energies $T_{\mathrm{min}}$ (**Figure S1**).
**pp. S3-S4:** DFT studies of the 1D nanoobject with D1 polycyclic regions: bond length alternation in carbon chains **(Figure S2)** and partial densities of states in the antiferromagnetic state **(Figure S3).**
**p. S5**: Description of the video files attached.

---

[1] Corresponding author. Tel. +7-909-967-2886 E-mail: popov-isan@mail.ru (Andrey Popov)

# Distributions of successful electron impacts over the energies transferred to hydrogen and carbon atoms

The minimal transferred energies $T_{min}$ used for hydrogen and carbon atoms are 1 and 5 eV, respectively. In 10 calculations with 80 keV energy for the chosen GNR structure, there are 2079 successful electron impacts producing structural changes. 675 (32%) out of these impacts are impacts on carbon atoms, 1404 are on hydrogen atoms (1362 (97%) resulting in knock-out of a hydrogen atom). The distributions of successful electron impacts over the energies transferred to hydrogen and carbon atoms show that most of the successful impacts occur at the energies that are significantly greater than the chosen values of $T_{min}$ (see Figure S1). Thus, almost all successful electron impacts are taken into account with such a choice of $T_{min}$.

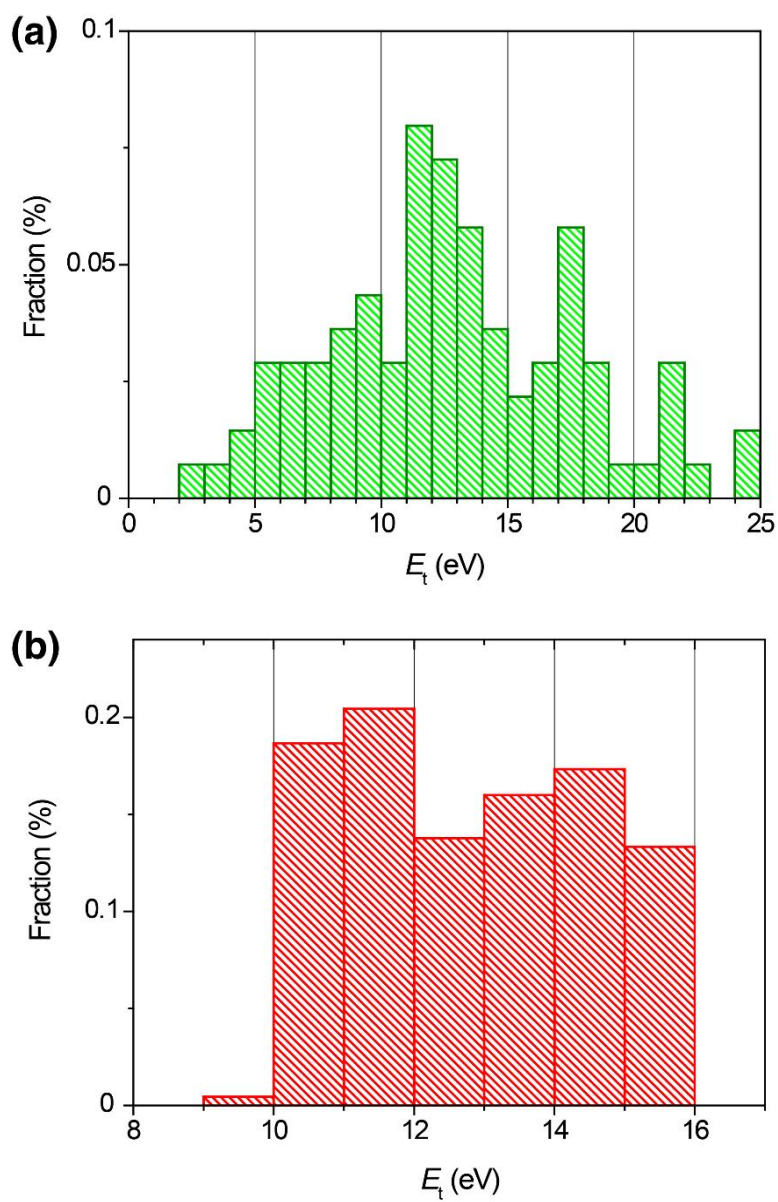

**Figure S1.** Calculated distributions of successful electron impacts over the energies transferred to hydrogen (a) and carbon (b) atoms. Each single bar shows the fraction of successful electron impacts (in %) corresponding to the given transferred energy $E_t$ (in eV) within the interval of 1 eV. The minimal transferred energies $T_{min}$ used for hydrogen and carbon atoms are 1 eV and 5 eV, respectively.

# DFT studies of the 1D nanoobject with D1 polycyclic regions

*Bond length alternation in carbon chains*

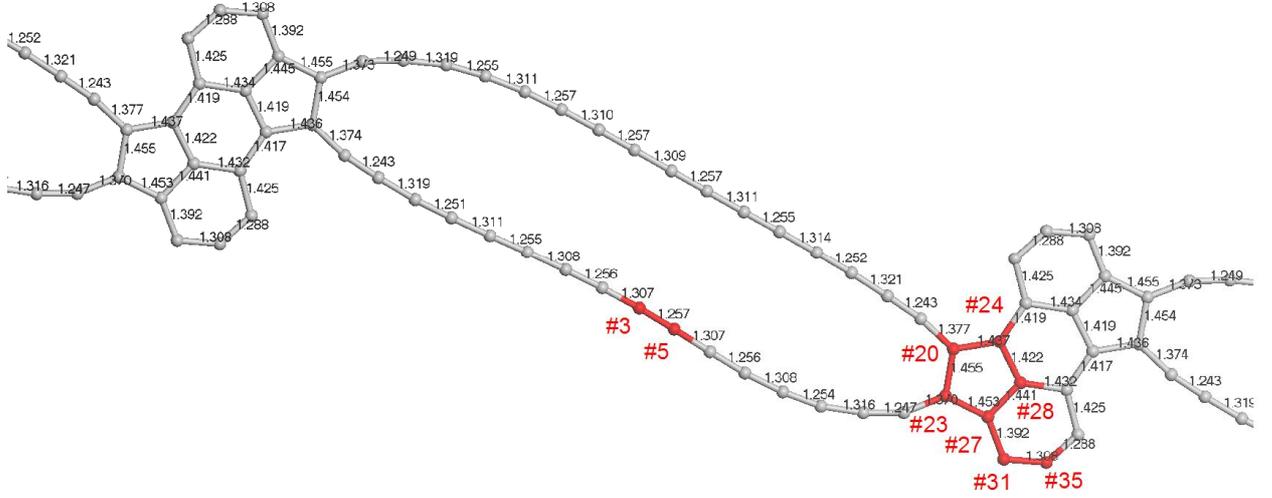

**Figure S2**. Geometrically optimized structure of the 1D nanoobject with D1 polycyclic regions obtained by the DFT calculations. The bond lengths (in Å) are indicated in grey. The atoms for which the partial densities of states are computed are shown in red.

The bond length alternation $\delta$ in carbon chains of the 1D nanoobject is computed according to papers [1,2] as

$$\delta = \frac{1}{2}\left| \frac{1}{n_e}\sum_{j=1}^{n_e}\left(d_{2j-1}+d_{n-(2j-1)}\right) - \frac{1}{n_0}\sum_{j=1}^{n_0}\left(d_{2j}+d_{n-2j}\right)\right|, \quad (S1)$$

where $d_i = |\vec{r}_i - \vec{r}_{i+1}|$, $n$ is the number of atoms in the chain, $n_e = (n+2)/4$, and $n_0 = n/4$ (integer part). The terminal bonds here are excluded.

In the 1D nanoobject under consideration, the chains are composed of 16 atoms each. Therefore, $n_e = n_0 = 4$. From the bond lengths shown in Figure S2, we get $\delta = 6.0$ pm and 5.8 pm for the upper and lower chains in this figure, respectively.

*Partial densities of states*

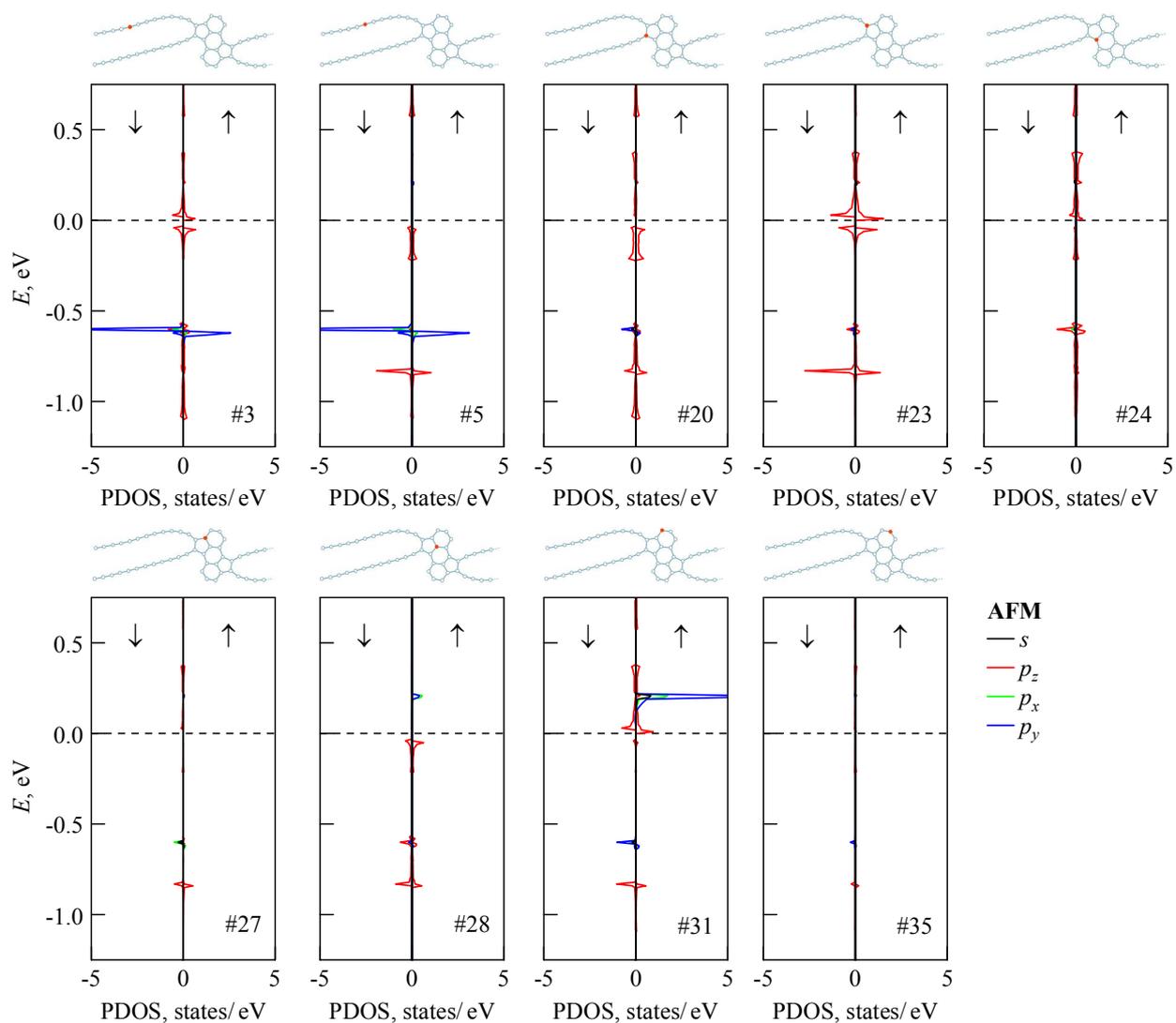

**Figure S3.** Calculated projected densities of states for selected atoms of the 1D nanoobject with D1 polycyclic regions (as indicated in the atomistic structures) in the antiferromagnetic state. The energies in eV are given relative to the Fermi level. The results for *s*, $p_z$, $p_x$ and $p_y$ orbitals are shown by black, red, green and blue lines, respectively.

**Description of the video files attached**

Video files "MD_SIMULATION_45keV.mp4", "MD_SIMULATION_55keV.mp4" and "MD_SIMULATION_80keV.mp4" show examples of the structure evolution observed in molecular dynamics simulations of the graphene nanoribbon transformation in HRTEM under electron irradiation with the electron kinetic energies of 45, 55 and 80 keV, respectively. In these video files, all carbon and hydrogen atoms are colored in grey and green, respectively. The total time in HRTEM (converted from the MD simulation time) is given at the bottom of the frame.

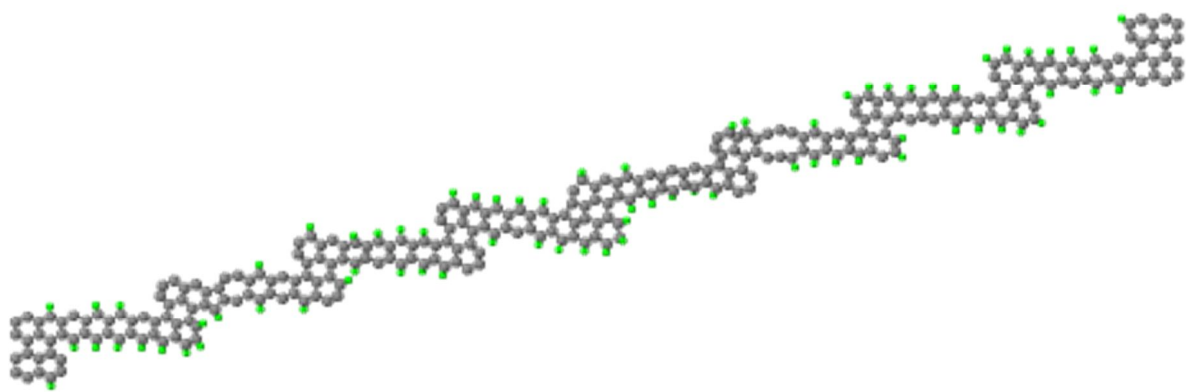